\documentclass[aps,prl,notitlepage,reprint,floatfix,amsmath,amssymb]{revtex4-2}

\usepackage{graphicx}
\usepackage{dcolumn}
\usepackage{bm}

\usepackage[separate-uncertainty = true]{siunitx}
\usepackage{ulem}

\begin{document}


\title{Evidence of localization effect on photoelectron transport induced by alloy disorder in nitride semiconductor compounds}
\author{Myl\`{e}ne Sauty$^{1}$}
\author{Nicolas M. S. Lopes$^{1}$}
\author{Jean-Philippe Banon$^{1}$}
\author{Yves Lassailly$^{1}$}
\author{Lucio Martinelli$^{1}$}
\author{Abdullah Alhassan$^{2}$}
\author{Shuji Nakamura$^{2}$}
\author{James S. Speck$^{2}$}
\author{Claude Weisbuch$^{1,2}$}
\author{Jacques Peretti$^{1}$}
\affiliation{$^1$Laboratoire de Physique de la Mati\`{e}re Condens\'{e}e, Ecole polytechnique, CNRS, Institut Polytechnique de Paris, 91120 Palaiseau, France}
\affiliation{$^2$Materials Department, University of California, Santa Barbara, California 93106, USA}

\date{\today}

\begin{abstract}
Near-bandgap photoemission spectroscopy experiments were performed on p-GaN and p-InGaN/GaN photocathodes activated to negative electron affinity. The photoemission quantum yield of the InGaN samples drops by more than one order of magnitude when the temperature is decreased while it remains constant on the GaN sample. This indicates a freezing of photoelectron transport in p-InGaN that we attribute to electron localization in the fluctuating potential induced by the alloy disorder. This interpretation is confirmed by the disappearence at low temperature of the peak in the photoemission spectrum that corresponds to the contribution of the photoelectrons relaxed at the bottom of the InGaN conduction band.
\end{abstract}

\maketitle

Alloying is a major tool to tune the electronic structure of semiconductors. However, except for very specific stoichiometry of particular compounds, alloys exhibit an unavoidable compositional disorder due to the random placement of the atoms on the crystal lattice sites. 
Disorder has been shown for decades to have a strong influence on the optical and electronic properties of semiconductors \cite{Shklovskii1984}. In particular, the intrinsic alloy disorder was proved to be responsible for the broadening of the absorption edge \cite{Bansal2007} as well as for exciton localization effects, as was observed in AlGaAs or GaAsN at low temperature \cite{Sturge1983, Saidi2002}. 

The case of nitride compounds obtained by alloying InN, GaN and AlN is particularly interesting. Indeed, the band gap of III-N ternary (and quaternary) alloys varies very strongly with composition.
Therefore, the alloy disorder induces potential fluctuations of tens to hundreds of meV on a scale of a few nanometers \cite{Wu2012, Filoche2017}. Such potential fluctuations are expected to induce localization effects even at room temperature. 

One-particle models, taking into account the intrinsic alloy disorder, indicate that the low energy states for holes are localized in nitride ternary compounds \cite{Schulz2015, DiVito2019, WatsonParris2011}, but the existence of localized states for electrons is still debated, both in 2D and 3D systems \cite{Schulz2015, DiVito2019, WatsonParris2011, Wu}. Extrinsic properties, like alloy clustering or quantum well thickness fluctuations, are proposed as causes of localized electron states \cite{DiVito2019, DiVito2020, Schulz2015}, but their contribution is questioned \cite{McMahon2020, Weisbuch2021}. In another respect, electron-hole Coulomb interaction, at the origin of the excitonic structures observed in absorption measurements \cite{David2019}, seems to be an important ingredient to lead to localized electron wavefunctions \cite{Tanner2018, ADavid_excitons}.

The effects of disorder on absorption ~\cite{Butte2018,Piccardo2017,David2019} and recombination \cite{Schoemig2004, Gotoh2006, Callsen2019, Solowan2013} in InGaN have been reported down to the intrinsic alloy disorder scale \cite{Hahn2018}. However, evidencing electron localization requires electron transport measurements as a function of temperature. Indeed, if low energy electronic states are localized in the fluctuating potential landscape, the transport of low energy carriers should not follow a drift-diffusion process but should occur via phonon-assisted hopping either between localized states or from localized states to higher energy delocalized states. These mechanisms are expected to strongly depend on temperature \cite{Tessler2009}. 

Characterizing carrier transport by usual electrical measurement techniques is a challenging issue in nitride ternary compounds. On the one hand, bulk materials are unavailable, so that measurements must be performed in heterostructures that incorporate a thin alloy layer. On the other hand, probing localized states requires low carrier density and therefore low doping level. Measuring the transport properties of thin, lowly doped alloy layers can hardly be achieved \cite{Schultes2013, Matioli2015} due to parasitic parallel current pathways in the neighboring thicker layers or substrate. Note that, the alloy disorder was shown to strongly reduce the mobility in high density 2D electron gases of an InGaN channel \cite{Sohi2018}. However, due to the very high carrier density, such experiment could not provide any information on the localization of electronic states.

Here, we report on the study of electron transport in p-doped InGaN/GaN heterostructures by near-band-gap photoemission spectroscopy. This technique relies on the activation of the p-type semiconductor surface to effective negative electron affinity (NEA) usually by deposition of a cesium monolayer. In the NEA situation, the conduction band minimum in the bulk semiconductor lies above the vacuum level so that photoelectrons excited with near-band-gap light can be emitted into vacuum \cite{Spicer1958, Scheer1965}. NEA photoemission spectroscopy is sensitive to the conduction band structure \cite{Drouhin1985, Lassailly1990, Peretti1991, Piccardo2014}. Furthermore, since the light absorption length is of the order of the electron diffusion length, it provides a unique spectroscopic access to electron transport processes \cite{Spicer1958, Peretti1993, Iveland2013, Ho2021}. This approach allowed us to probe minority electron transport in thin InGaN layers at very low electron concentration, without the limitations of usual electrical transport measurements. We observe the freezing of low-energy photoelectron transport at low temperature in the disordered InGaN alloy. This shows that, even in a 3D system, the low energy electron states are localized, and that electron transport occurs by thermally assisted processes.

\begin{figure}
\includegraphics[scale=0.25]{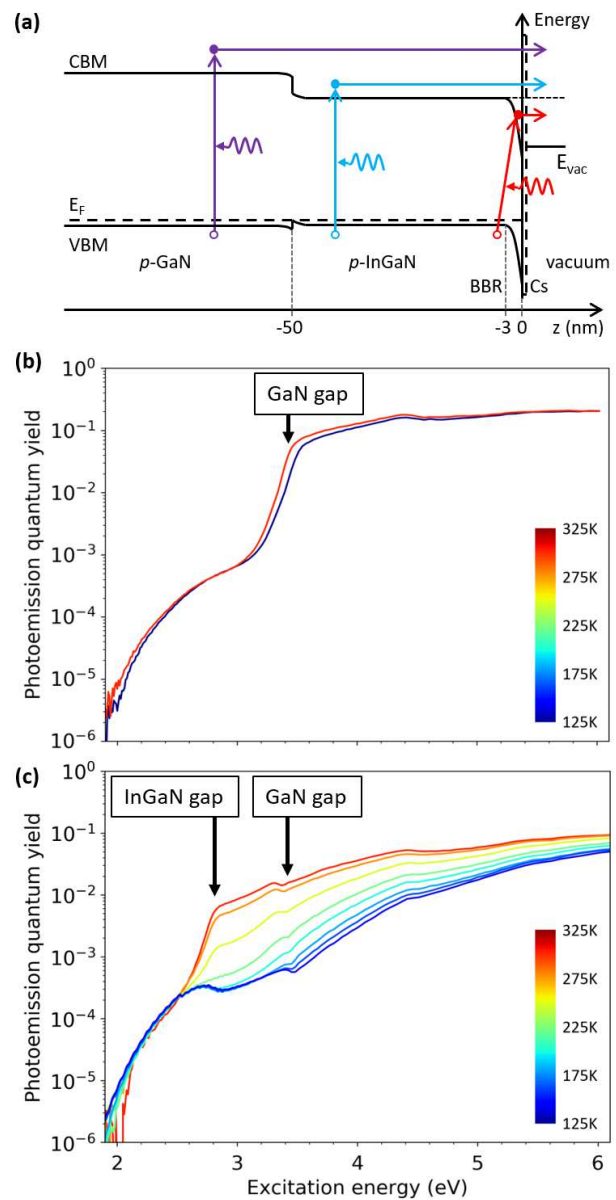}
\caption{\label{fig:QY}(a) Calculated bandstructure in real space of the InGaN/GaN heterostructures showing the conduction band minimum (CBM) and valence band maximum (VBM). The different photoemission processes are indicated by colored arrows. With below band gap excitation, electrons can be photoemitted from the near-surface BBR. With above bandgap excitation, electrons photoexcited in the InGaN layer or in the underlying GaN layer can be photoemitted. (b) and (c) Excitation spectra of the QY measured on the p-GaN and In$_{0.15}$Ga$_{0.85}$N/GaN samples, respectively. The positions of In$_{0.15}$Ga$_{0.85}$N and GaN gaps at 300~K are indicated. The colour of each spectrum indicates the temperature at which it was acquired according to the color scale shown in inset.}
\end{figure}

The studied c-plane InGaN/GaN heterostructures were grown by MOCVD on a sapphire substrate. 
They consist of a top 50~nm p-doped InGaN layer, a 75~nm p-GaN layer, a 2~$\mu$m n-GaN layer grown on a GaN buffer layer deposited on a sapphire substrate. The InGaN and GaN p-layers are Mg doped at concentrations of $\SI{1e19}{}$ and $\SI{6e19}{\cm^{-3}}$, respectively, with overdoping of the top $\SI{10}{\nm}$ of InGaN. The calculated band diagram in real space is shown in Fig.~\ref{fig:QY}(a). Assuming that the Fermi level E$\mathrm{_{F}}$ is pinned near the InGaN mid-gap at the sample surface, the width of the band bending region (BBR) close to the surface is of a few nm, i.e. much shorter than both the light absorption length and the InGaN layer thickness. To reveal the effects of alloy disorder on the photoemission process, two InGaN/GaN samples have been studied with respectively 5\% and 15\% In content. In addition, control measurements have been performed on a GaN sample, consisting of a 200 nm-thick p-doped GaN layer, with a Mg concentration of $\SI{5e19}{\cm^{-3}}$, and surface overdoping. 

The samples were chemically cleaned consecutively with piranha and HCl-isopropanol solutions \cite{Tereshchenko2004}. They were then introduced into the UHV chamber, with base pressure in the low $10^{-11}$~mbar, annealed for ten minutes at 350$^{\circ}$C, and immediately after, activated to NEA by cesium deposition. Activation was controlled by monitoring the photoemission current under excitation with near-band-gap light. With this procedure, the work function was typically reduced to $\sim$1.6 eV which corresponds to an effective NEA of about -1.8 eV on GaN (-1.6 eV and -1.2 eV on the InGaN samples). This NEA state was stable for several hours.
	
The photoemission quantum yield (QY), i.e., the number of emitted electrons per incident photon, was measured as a function of the photon energy $h\nu$, for different sample temperatures. 
The excitation wavelength was scanned from 690~nm (1.8~eV) to 200~nm (6.2~eV) with an output bandwidth of 5~nm. In this spectral range, the illumination setup delivers an output power density which varies monotonously from $100$ to $\SI{1}{\micro \watt \cm^{-2}}$. The corresponding estimated photocarrier concentration is less than $10^{11}$~$\SI{}{\cm^{-3}}$.

The excitation spectra of the QY recorded on the GaN sample at 125~K and 300~K are shown in Fig.~\ref{fig:QY}(b). A below bandgap photoemission regime is observed as usually in NEA semiconductor photocathodes \cite{Pakhnevich2004, Uchiyama2005}. We attribute this regime to Franz-Keldysh processes in the near-surface BBR as shown schematically in Fig.~\ref{fig:QY}(a) \cite{BBR}. When approaching the GaN bandgap energy, the QY increases abruptly by almost two orders of magnitude. Then, for above bandgap excitation the QY slowly increases. Near 4.5 eV, a kink is observed, probably related to photoelectron transferred in the first side valley of the conduction band \cite{Piccardo2014}. Above 5 eV excitation energy, the QY reaches 0.2, a value comparable to already reported ones \cite{Tereshchenko2004, Pakhnevich2004} but well below the performances of industry-optimized photocathodes \cite{Uchiyama2005}. Just above the GaN bandgap, the QY is 0.07. Using the Spicer's model for NEA photoemission \cite{Spicer1958}, with a light absorption length of 100 nm and an extraction coefficient of 0.2 (consistently with the maximum QY obtained at high excitation energy), we can estimate that the minority electron diffusion length is of about 50 nm in GaN, in agreement with already reported values \cite{Kumakura2005, Kumakura2007}. When decreasing the temperature to 125~K, the QY excitation spectrum of the p-GaN remains nearly unchanged, except for a slight blue shift around 3.4~eV due to the bandgap increase.

The QY excitation spectra measured on the In$_{0.15}$Ga$_{0.85}$N sample at different temperatures, are plotted in Fig.~\ref{fig:QY}(c). The below bandgap photoemission regime originating from the near-surface BBR is also observed. Then, the QY excitation spectra exhibit features at characteristic energies of the sample band structure. First, a significant increase, by more than one order of magnitude, is observed at room temperature at the InGaN band gap. Second, kinks show up at 3.4 eV and 4.5 eV which originate from electrons excited in the underlying p-GaN layer. This demonstrates that NEA photoemission is sensitive to electrons transported all through the 50~nm-thick p-InGaN layer.

\begin{figure}[b]
\includegraphics[scale=0.23]{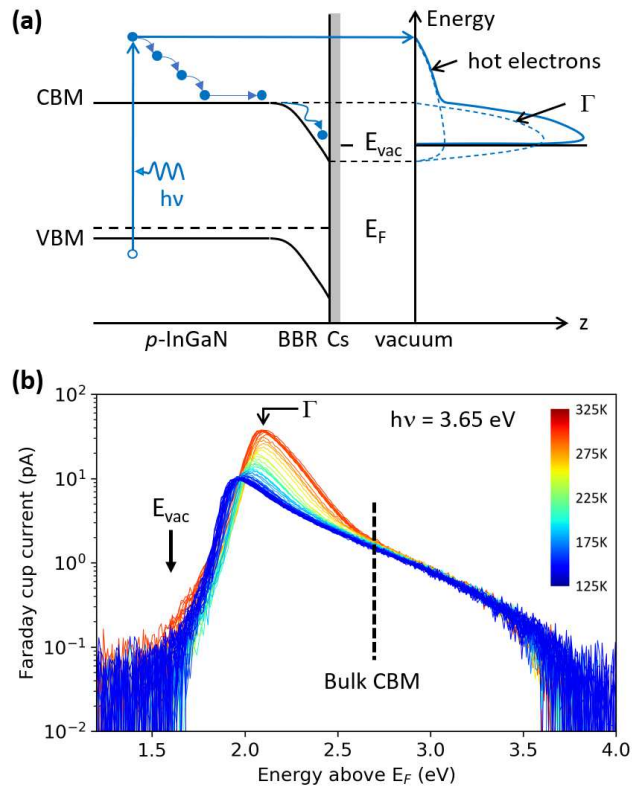}
\caption{\label{fig:EDC} (a) Schematics of the photoemission process. The EDC is delimited at low energy by the vacuum level E$\mathrm{_{vac}}$ and at high energy by the final state of the optical transition from E$\mathrm{_{F}}$. A characteristic low-energy peak labeled $\Gamma$ is formed by photoelectrons which accumulate at the bottom of the conduction band in the bulk semiconductor and partially relax their energy in the BBR before emission. (b) EDCs measured at different temperatures between 140~K and 300~K on p-In$_{0.15}$Ga$_{0.85}$N, with $h\nu=3.65$~eV. The colour of each spectrum indicates the temperature at which it was acquired according to the colour scale shown in inset.}
\end{figure}

The striking difference between the QY excitation spectra of In$_{0.15}$Ga$_{0.85}$N and GaN lies in their dependence on temperature. When the temperature decreases, the QY of the p-In$_{0.15}$Ga$_{0.85}$N/GaN structure drops by more than one order of magnitude for excitation energies above the In$_{0.15}$Ga$_{0.85}$N bandgap, while the QY of the p-GaN sample is nearly unchanged over the whole excitation energy range. A decrease in the QY when decreasing the temperature can be due either to the freezing of electron transport or to an increase of the vacuum level caused by either cryogenic trapping of contaminants or surface photovoltage. 

In order to discriminate between these different effects, we have measured the energy distribution curve (EDC) of the photoemitted electrons with an electron spectrometer specifically designed for low-energy operation \cite{Drouhin1986}. The EDC lineshape is determined by the photon energy, the semiconductor band structure and the transport processes between excitation and emission, as schematized in Fig.~\ref{fig:EDC}(a). Mainly two contributions are expected. Electrons which accumulate in the $\Gamma$ valley at the bottom of the conduction band in the bulk and lose part of their energy in the BBR before emission, give rise to an intense low energy peak (labeled $\Gamma$) with a high energy threshold pointing at the conduction band minimum (CBM) in the bulk. Electrons excited in the bulk and in the BBR which have only partially relaxed their energy before emission (without accumulation at the bulk CBM) give rise to a broad but weak hot electron distribution which extends from the vacuum level up to the final state energy of the optical transition in the bulk. 

The EDCs of the photoemitted electrons obtained for $h\nu=3.65$~eV at different temperatures on the In$_{0.15}$Ga$_{0.85}$N sample are plotted in Fig.~\ref{fig:EDC}(b) in logarithmic scale. The low-energy threshold of the EDCs corresponds to the vacuum level position. It lies around 1.6~eV above $E_F$, which corresponds to a NEA of about -1.2~eV.
At room temperature, the EDC exhibits an intense low-energy peak with a high energy threshold at 2.7~eV above $E_F$. It corresponds to the contribution from electrons accumulated in the bulk CBM. Considering that the In$_{0.15}$Ga$_{0.85}$N gap and the dopant activation energy are respectively 2.8~eV and 80~meV \cite{Kumakura2000}, the high energy threshold of this contribution coincides with the bulk CBM in InGaN. In addition, the EDC exhibits a hot-electron contribution of much lower intensity, which extends well above the bulk CBM.

When decreasing the temperature, the $\Gamma$ contribution almost completely disappears while the hot-electron contribution remains unchanged. The vacuum level position also does not change, which indicates that there is neither deterioration of the NEA activation nor surface photovoltage effects that could lead to a decrease in the QY with decreasing temperature.
The disappearance of the $\Gamma$ peak at low temperature shows that the transport of electrons relaxed in the low energy states of the conduction band in bulk p-InGaN is frozen at low temperature. This is at the origin of the observed drop in the QY with decreasing temperature \footnote{Note that for EDC measurements, the incident power density was higher than for the QY measurements. This is the reason why the drop in the EDCs integrated intensity is smaller than in the QY measurements of Fig. \ref{fig:QY}}. 

\begin{figure}[t]
\includegraphics[scale=0.31]{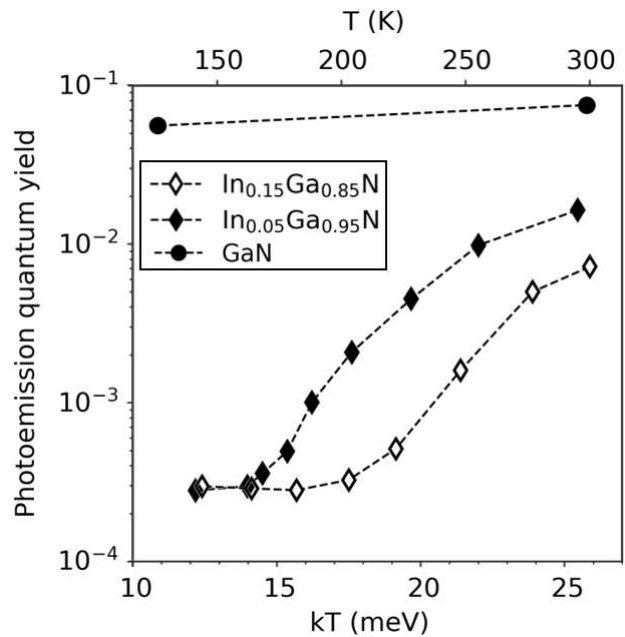}
\caption{\label{fig:CutsQY} Variation of the QY versus temperature, at a fixed, above-band-gap excitation energy: 2.9, 3.3 and 3.5~eV, respectively for In$_{0.15}$Ga$_{0.85}$N, In$_{0.05}$Ga$_{0.95}$N and GaN.}
\end{figure}

The same experiment was performed on an InGaN sample with 5\% In content. In Fig. \ref{fig:CutsQY} are plotted the variations of the QY measured on In$_{0.15}$Ga$_{0.85}$N, In$_{0.05}$Ga$_{0.95}$N and GaN, just above their respective band gap, as a function of temperature. 
In both InGaN samples, when the temperature decreases, the QY abruptly drops down to a plateau value which corresponds to the integrated intensity of the hot electron contribution. This drop occurs at lower temperature for the InGaN alloy with 5\% In content. 
This freezing of electron transport at low temperature shows that transport occurs through thermally assisted processes. 
This strongly supports the fact that low energy electrons are localized in the disordered potential of the InGaN alloy.
Then, for lower disorder, i.e. for lower In content, the freezing temperature of thermally assisted transport is lower.

As already mentioned, the existence of electron localized states induced by intrinsic compositional disorder in InGaN alloys is widely disputed on the basis of one-particle simulations \cite{Schulz2015, DiVito2019}. However, additional effects can be considered.

Weak localization could explain the freeze out of low energy electrons at low temperature \cite{Akkermans2007}. Indeed, the scattering of the electronic wave packets on the disordered potential can strongly reduce their diffusion length and induce an effective localization of low energy electrons. 
This effective localization depends on temperature since, when the temperature is high enough, low energy electrons can access completely delocalized states. However, the calculation of the plane waves elastic scattering rate on the disordered potential in the effective mass approximation indicates that the mean free path between two scattering events is of a few tens of nm, leading to an effective localization length certainly larger than the InGaN layer thickness \cite{Banon}. It is therefore very unlikely that weak localization plays a significant role.   

In contrast, there is strong indications that alloy disorder induces localized hole states. 
Moreover, it has been shown both numerically \cite{ADavid_excitons, Tanner2018} and experimentally by the observation of exciton peaks in absorption \cite{David2019} that the Coulomb interaction between electrons and holes in disordered InGaN is of the order of a few tens of meV, and could lead to the localization of the electron wavefunction. 
In the p-InGaN samples studied here, the ionized acceptor density is about $\SI{5e17}{\cm^{-3}}$ at 300~K and  $\SI{1e16}{\cm^{-3}}$ at 140~K \cite{Kumakura2000}. The density of photoexcited holes is negligible. 
The typical size of hole localization subregions being of about 5~nm according to simulations \cite{Filoche2017, Piccardo2017}, the density of localization subregions is of about $\SI{8e18}{\cm^{-3}}$. 
This means that holes are all localized, with an average distance between two holes of about 15~nm at 300~K and 50~nm at 140~K. During the photoemission process, most of the excited electrons relax by phonon emission towards the bottom of the conduction band, i.e., to the lowest energy states they can find, and form the $\Gamma$ peak observed on the room temperature EDCs. Since the electrons escape probability at the surface is small (at room temperature, the QY reaches at best 0.1 and 0.2 in the InGaN samples with 15\% and 5\% In content, respectively), it is probable that electrons explore a large enough area to find a localized hole to which they bind. At room temperature, thermal excitation would allow the photoelectrons relaxed in these localized states to access delocalized states. 
Another possibility would be for Coulomb bound electron-hole pairs to hop between localization regions that are close enough in energy \cite{ADavid_excitons}. 
These processes are much less efficient at low temperature and transport of low-energy electrons bound to localized holes would be frozen. Fig.~\ref{fig:CutsQY} shows that the transport of electrons relaxed at the bottom of the InGaN conduction band is frozen at 200~K and 160~K for In$_{0.15}$Ga$_{0.85}$N and In$_{0.05}$Ga$_{0.95}$N, respectively, which correspond to critical thermal energies of about 17 meV and 14 meV, comparable with characteristic electron-hole binding energies in InGaN \cite{ADavid_excitons}.

In conclusion, we have performed near-bandgap photoemission on p-type GaN and InGaN/GaN heterostructures, activated to NEA. In InGaN with both 15\% and 5\% In content, the QY drops dramatically when decreasing the temperature, due to the freezing of the transport of low energy photoelectrons. This indicates that low energy electron states are localized in the disordered potential of the InGaN alloys and that electron transport occurs through thermally assisted processes. The temperature at which transport is frozen decreases with decreasing In content, as expected since compositional disorder effects should decrease with decreasing In content. It should be noted that these results are obtained for 3D systems, where localization effects are expected to be weaker and transport easier than in 2D quantum wells. These results contradict the theoretical predictions obtained from one-particle models. However, although the hole density is rather small in the studied p-type materials, especially at low temperature, it might be that the electron-hole Coulomb interaction plays a significant role in the electron localization.

We thank Aurélien David for fruitful discussions and Yuh-Renn Wu for fruitful discussions and access to the drift-diffusion charge control (DDCC) solver. This work was supported by the French National Research Agency (ANR, ELENID grant No. ANR-17-CE24-0040-01), by the Simons Foundation (Grants Nos. 601952 J.S.S., 601954 C.W., and 601944 J.-P.B.), the National Science Foundation (NSF) RAISE program (Grant No. DMS-1839077), ARPA-E, U.S. Department of Energy (program DE-EE0007096), the UCSB Solid State Lighting and Energy Electronics Center, and KACST-KAUST-UCSB Solid State Lighting Program (SSLP).


%

\end{document}